\documentclass[conference]{IEEEtran}

\usepackage[utf8]{inputenc}

\usepackage{graphicx}
\graphicspath{{assets/}}
\usepackage{glossaries}
\usepackage{listings}
\usepackage{pifont}
\usepackage{textgreek}
\usepackage{soul,color} 
\usepackage{threeparttable}
\newcommand{\cmark}{\ding{51}}

\newacronym{3GPP}{3GPP}{3rd Generation Partnership Project}
\newacronym{5G}{5G}{Fifth Generation}
\newacronym{API}{API}{application programming interface}
\newacronym{AR}{AR}{augmented reality}
\newacronym{CapEx}{CapEx}{capital expenditure}
\newacronym{DevOps}{DevOps}{Development Operations}
\newacronym{DSP}{DSP}{digital signal processor}
\newacronym{E2E}{E2E}{end-to-end}
\newacronym{LTE}{LTE}{Long Term Evolution}
\newacronym{EC}{EC}{edge computing}
\newacronym{MEC}{MEC}{multi-access edge computing}
\newacronym{MNO}{MNO}{mobile network operator}
\newacronym{NFV}{NFV}{network function virtualization}
\newacronym{NIC}{NIC}{network interface controller}
\newacronym{OpEx}{OpEx}{operating expenditure}
\newacronym{OSS}{OSS}{open-source software}
\newacronym{OTA}{OTA}{over-the-air}
\newacronym{QoS}{QoS}{quality of service}
\newacronym{RAN}{RAN}{radio access network}
\newacronym{SDN}{SDN}{software defined network}
\newacronym{SIM}{SIM}{subscriber identity module}
\newacronym{UE}{UE}{user equipment}
\newacronym{VOR}{VOR}{Vestibulo-Ocular Reflex}
\newacronym{VR}{VR}{virtual reality}
\newacronym{MR}{MR}{mixed reality}
\newacronym{SDR}{SDR}{software-defined radio}
\newacronym{PaaS}{PaaS}{Platform-as-a-Service}
\newacronym{HMD}{HMD}{head-mounted display}
\newacronym{IoT}{IoT}{Internet of Things}
\newacronym{GPU}{GPU}{graphics processing unit}
\newacronym{EPC}{EPC}{evolved packet core}
\newacronym{GPGPU}{GPGPU}{general-purpose computing on GPUs}
\newacronym{FPS}{FPS}{frames per second}
\newacronym{IR}{IR}{intermediate representation}
\newacronym{0-RTT}{0-RTT}{zero round trip time resumption}
\newacronym{RPC}{RPC}{remote procedure call}
\newacronym{URLLC}{URLLC}{ultra-reliable low-latency communication}
\newacronym{CP}{CP}{cyclic prefix}
\newacronym{OFDMA}{OFDMA}{Orthogonal Frequency Division Multiple Access}
\newacronym{OFDM}{OFDM}{Orthogonal Frequency Division Multiplexing}
\newacronym{HARQ}{HARQ}{Hybrid Automated Repeat Request}
\newacronym{TTI}{TTI}{Transmission Time Interval}
\newacronym{KPI}{KPI}{Key Performance Indicator}
\newacronym{FFT}{FFT}{fast Fourier transform}
\newacronym{PAPR}{PAPR}{peak-to-average power ratio}
\newacronym{SoC}{SoC}{system on chip}
\newacronym{GLSL}{GLSL}{OpenGL shading language}
\newacronym{COTS}{COTS}{commercial off-the-shelf}
\newacronym{UDP}{UDP}{user datagram protocol}
\newacronym{QUIC}{QUIC}{quick UDP internet connections}
\newacronym{ETSI}{ETSI}{European Telecommunications Standards Institute}
\newacronym{DSL}{DSL}{domain-specific language}
\newacronym{MIR}{MIR}{mid-level intermediate representation}

\title{Interoperable GPU Kernels as Latency Improver for MEC}
       
\author{\thanks{This is the accepted version of the work. The final version will be published at the 2nd 6G Wireless Summit (6G SUMMIT), March 17-20, Levi, Finland, 2020.}
\IEEEauthorblockN{Juuso Haavisto, and Jukka Riekki}
\IEEEauthorblockA{
   Center for Ubiquitous Computing, University of Oulu\\
   Email: \{\small \tt first.last\}@oulu.fi}
}
\IEEEoverridecommandlockouts

\begin{document}

\maketitle

\begin{abstract}

Mixed reality (MR) applications are expected to become common when 5G goes mainstream. However, the latency requirements are challenging to meet due to the resources required by video-based remoting of graphics, that is, decoding video codecs. We propose an approach towards tackling this challenge: a client-server implementation for transacting intermediate representation (IR) between a mobile UE and a MEC server instead of video codecs and this way avoiding video decoding. 
We demonstrate the ability to address latency bottlenecks on edge computing workloads that transact graphics. We select SPIR-V compatible GPU kernels as the intermediate representation.
Our approach requires know-how in GPU architecture and GPU domain-specific languages (DSLs), but compared to video-based edge graphics, it decreases UE device delay by sevenfold. Further, we find that due to low cold-start times on both UEs and MEC servers, application migration can happen in milliseconds.
We imply that graphics-based location-aware applications, such as MR, can benefit from this kind of approach.
    
\end{abstract}

\section{Introduction}

\gls{5G} is expected to introduce new \glspl{UE}, which use the low-latency aspects of the new wireless standard. In literature, typical examples include hands-free, eyes-on user interfaces such as \gls{MR} devices. \gls{MR} devices combine the pass-through capabilities of \glspl{HMD} to project holographic content to the users' visual periphery, as seen either through the \gls{VR} \glspl{HMD} front-facing cameras or through translucent screens of \gls{AR} \glspl{HMD}. At present, \gls{MR} experiences are achievable via \gls{COTS} hardware when utilizing hardware meant for gaming or other graphics-heavy applications. 

However, the envisioned 5G devices have small form factors and hence cannot use such hardware. This challenge can be tackled with \gls{MEC} and its new low-latency edge computing capabilities. In \gls{MEC}, computation offloading is executed on a \gls{MNO} server in the immediate periphery of the serving base station. Leveraging such \gls{MNO}-maintained edge computing infrastructure paves the way for the practical guarantees via which the \gls{5G} devices could rely on the cellular infrastructure for computation. Coincidentally, this would result in a way to make the \glspl{UE} small. Yet, despite such MEC-dependent hardware being often envisioned, empirical studies in realizing such devices are rare, especially considering latency deadlines. Due to the lack of such empirical research, we hereby present such in this paper: we demonstrate an approach that addresses the \gls{E2E} latency between the \gls{MEC} server and the mobile \gls{UE}.

For MR applications, MECs can be used to create graphcis for UEs. In general, our method to reduce latency questions whether \gls{MEC} applications should remote graphics via video codecs, as the current cloud computing equivalents do \cite{suznjevic2016analysis}. While these video streaming services, such as Google Stadia, work with home-network fiber connections, previous studies \cite{kamarainen2017measurement} show mobile \glspl{UE} having a latency problem with video codec decoding. Here, the \gls{UE} must decode and render video frames while maintaining a low-latency and reliable cellular connection. According to Kämäräinen et al. \cite{kamarainen2017measurement}, the decoding process takes longer than what is required to achieve a real-time (i.e., 60 \gls{FPS}) computation offloading. This implies latency bottlenecks for the envisioned \gls{5G} applications. 

Our contribution is removing this bottleneck with \gls{IR}. We leverage the interoperability of the Vulkan graphics \gls{API} and the SPIR-V \gls{IR} via a \gls{RPC} interface between \gls{UE} and \gls{MEC} server. Applications can be run at 60 \gls{FPS} on \glspl{UE} with a lightweight form factor because real-time graphics can be offloaded to \glspl{MEC} without the need for decoding video frames. The cold-start times of the \gls{GPU} kernels on both UE and MEC are on a millisecond scale, which enables migrating applications in real-time, possibly even during a single frame refresh.


The rest of the paper is structured as follows: Section §\ref{ch:bg} introduces \gls{5G} networks and \gls{GPU} \glspl{DSL}. Section §\ref{ch:sf} reasons our chose of \gls{GPU} \gls{DSL} and \gls{API}. Section §\ref{ch:ir} presents the implementation and results, and §\ref{ch:di} and §\ref{ch:co} the discussion and conclusions, respectively.

\section{Background}
\label{ch:bg}

\subsection{Latency in Multi-Access Edge Computing}

\gls{5G} cellular networks include the \gls{ETSI} standardized \gls{MEC} paradigm, which places orchestrated cloud computing resources within the local \gls{RAN}, hosted by a \gls{MNO}. As such, \gls{MEC} introduces a new way to optimize service latency: \glspl{MNO} may design networks in which third-party middleboxes are avoided, and network hops reduced. These kinds of topology optimizations facilitate guarantees for latency, jitter, and throughput, as network requests need not traverse beyond the local area network of the \gls{RAN}. Further, knowledge about the physical network design can be leveraged for purpose-built communication stacks and other optimizations. For example, \gls{UDP}-based protocols like \gls{QUIC} perform better in such networks, as less port punch-holing is required for the lack of legacy network equipment suppressing packet flow and delivery.

In the \gls{MEC} paradigm, latency can be decomposed into the following general categories: (1) \textit{access delay}, which is physics bound, (2) \textit{device delay}, which concerns the \gls{UE}, and (3) \textit{server delay}, which is the \gls{RAN} infrastructure. Together, these factors form the \gls{E2E} latency. Below, we further define the particularities of each category.

\subsubsection{Access delay}

In \glspl{RAN}, packet delivery is specified to happen within $0.5$~ms for downlink and $0.5$~ms for uplink. Here, the latency is the time it takes to deliver an application layer packet from the radio protocol layer $2/3$ SDU ingress point to the radio protocol layer $2/3$ SDU egress point via the radio interface in both uplink and downlink directions. According to specifications \cite{etsi}, this assumes error-free conditions, and utilizing all assignable radio resources for the corresponding link direction. As the $0.5$~ms latency deadline consists of a physics-bound air interface rather than software-bound limits, it could be considered the baseline latency for any higher-order communication, such as anything happening via the \gls{MEC}. 

\subsubsection{Device delay}

For offloading graphical end-user applications, some physical constraints must be considered, such as \gls{VOR} with \gls{MR} applications. Here, applications regarding the \gls{VOR} need a screen refresh rate of $120$~Hz, which leaves $1000\mathrm{ms}\div120-1\mathrm{ms}=7.33\mathrm{ms}$ of overhead to spend on the \textit{device delay}, i.e., for processing on the UE. This time-window must then contain the (1) input, (2) rendering, (3) display, (4) synchronization, and (5) frame-rate-induced delay \cite{kamarainen2017measurement, jacobs1997managing}. As such, this part of the roundtrip is arguably the most challenging of the three. Coincidentally, the device delay is the main focus of our study.

\subsubsection{Server delay}

What is left from \textit{device delay} (i.e. from $7.33$~ms) can then be spent on \textit{server delay}, which occurs in the wired \gls{RAN} backbone. We define this latency as the time it takes for the roundtrip packet delivery from the base station, first to the \gls{EPC}, and finally to the \gls{MEC} server. With \gls{MEC}, the software in the environment is specified to be virtualized and orchestrated, hence, the virtualization approach is an important piece to achieve low-latency. For transacting low-latency computation or graphics, the \gls{MEC} infrastructure should also account for low-latency cold-start times. The cold-start times are relevant in case the quality of the cellular connectivity decreases, and the \gls{UE} would need to start the same program on its hardware. The same applies for migrating software between the \gls{MEC} servers, and is useful in radio-handovers: when the cellular network decides to move a \gls{UE} from a serving base station to another, the MEC orchestrator needs to either pre-emptively or proactively move the current session to a \gls{MEC} server closer to the new base station to optimize latency. In our previous study \cite{okwuibe2019orchestrating}, we observed it takes seconds to proactively move a container-based session managed by Kubernetes from a MEC server to another. Yet, for high-quality end-user experiences, this migration time would most certainly need to be fast enough to be imperceptible.

\subsection{GPU Programming}
\label{s:gpgpu}

\begin{table} 
\caption{Comparison of GPU APIs.}
\begin{center}
\begin{tabular}{ c c c c }
 & Compiler & Shader & Initial \\
 & Target & Language & Release \\
 \hline
 OpenGL &  & \cmark & 1992 \\
 OpenCL & \cmark & \cmark & 2009 \\  
 Vulkan & & \cmark & 2016 \\
 OpenMP & \cmark & & 1997  \\ 
 CUDA & \cmark & & 2007 \\ 
 OpenACC & \cmark  & & 2011 \\
 AMD GCN & \cmark  & & 2011 \\ 
 C++ AMP & \cmark  & & 2012 
\end{tabular}
\end{center}

\label{fig:apis}
\end{table}

\begin{table*} 
\begin{minipage}{\textwidth}
\begin{center}
\begin{threeparttable}[b]
\caption{Shader language and platform support for GPU APIs.}
\label{fig:support}
\begin{tabular}{ c | c | c | c | c c c c c c c }
 Supports & GLSL & OpenCL C & SPIR-V & Nvidia & AMD & Android & iOS & CPUs & FPGAs & DSPs \\
 \hline
 OpenGL & \cmark & & \cmark & \cmark & \cmark & \cmark \\ 
 OpenCL & \cmark & \cmark & \cmark & \cmark & \cmark & & & \cmark & \cmark & \cmark \\ 
 Vulkan & \cmark\tnote{a} & \cmark\tnote{b} & \cmark & \cmark & \cmark & \cmark & \cmark\tnote{c}  \\ 
 \hline
 Initial Release & 2004 & 2009 & 2014
\end{tabular}
\begin{tablenotes}
\item [a] With glslang compiler
\item [b] With clspv compiler
\item [c] With MoltenVK
\end{tablenotes}
\end{threeparttable}
\end{center}
\end{minipage}
\end{table*}

\gls{GPU} shaders are, in general, programs containing instructions that produce output to a framebuffer. The framebuffer is then used to render graphics onto a monitor. \glspl{GPU} cores can also be used for other parallel workloads than graphics, e.g., for matrix multiplications. This so-called \gls{GPGPU} paradigm has recently become a powerful tool in accelerating various parallel applications, e.g., machine learning and cryptocurrency mining. In general, to utilize the \gls{GPU} for graphical or general computation, the \gls{GPU} has to be given some form of \glspl{IR}. Table \ref{fig:apis} presents \gls{GPU} \glspl{API} which consume \glspl{IR}, and their supported programming approaches. Here, kernels, which are compilation targets, (1) cannot produce video output, (2) are platform- or \gls{API}-specific, and (3) presented as part of an existing programming language. E.g., OpenCL C uses compiler extension pragmas to translate subset of C into \gls{GPU} instructions. Such instructions are then consumable via the OpenCL \gls{API} and the platforms that support it (see: Table \ref{fig:support}). Shader languages, on the other hand, can do both graphics production and \gls{GPGPU}. With shader languages, e.g., GLSL, the \gls{IR} is platform- and \gls{API}-independent, but computations must be expressed in graphics terms like vertices, textures, fragments, and blending. As a result, the heterogeneity of traditional programming languages is lost, and shaders must be coded in their own language. Such a graphic-centric programming method is unwelcoming to programmers.

To reduce learning-curve, \glspl{DSL} have been introduced to make \glspl{GPU} more accessible. Some of these \gls{DSL}'s, such as Futhark \cite{henriksen2017futhark} are complete programming languages, while others, e.g., RLSL \cite{rlsl}, rely on existing programming environments and languages. Common to both, the projects aim to simplify \gls{GPU} programming by hiding away chores like shader language selection, generation, compilation, and the communication between the GPU and the CPU inside the language compiler.

In this work, we use shaders written in GLSL. For refined access to the graphics pipeline, we leverage the Vulkan GPU API. In our tests, the Vulkan API wrapper compiles the GLSL shaders to SPIR-V during runtime. This is required as Vulkan only supports SPIR-V \gls{IR} natively.

\section{System Framework}
\label{ch:sf}

\subsection{Prior Art}

Vulkan and SPIR-V are supported by some \gls{GPGPU} \glspl{DSL}, such as the aforementioned RLSL \cite{rlsl} and Futhark \cite{henriksen2017futhark, futharkVulkan}. RLSL is based on Rust programming language and extends its \gls{MIR}, whereas Futhark is a standalone language. Further, studies such as \cite{mammeri2018vcomputebench, DBLP:conf/hpca/WuBCCCDHIJJLLLQ19} remark Vulkan as promising cross-platform \gls{GPGPU} computing. Yet, none of these approaches address the idea of using kernel interoperability to transact graphics or reduce cold-start times in the area of telecommunications and the \gls{MEC}.

\subsection{Study focus}

In this study, we focus on interoperability and, with that, on \gls{E2E} latency reduction for graphics. We consider reducing the latency of general-purpose computing as well, hence we mean with interoperability that any edge computing workload done on-device on the \gls{UE} should be possible on the \gls{MEC} as-is, with the same source-code, and vice-versa. Hence, every platform-specific compilation target is out of the question. Per Table \ref{fig:apis}, the \glspl{API} are limited to OpenGL, OpenCL, and Vulkan. As Table \ref{fig:support} demonstrates, Vulkan is the only \gls{GPU} \gls{API} with native support for desktop \glspl{GPU} and mobile platforms such as Android and iOS. Thus with Vulkan, the \gls{MEC} could use Nvidia's and AMD's \glspl{GPU}, while having a single \gls{IR} compatible with the \glspl{UE}, which might use, e.g., Mali \glspl{GPU}. In other words, using Vulkan fulfills our focus on platform interoperability. To achieve this, Vulkan only supports SPIR-V natively. However, support of GLSL and OpenCL C can be achieved through separate compilers, which then produce SPIR-V IR. As GLSL can represent both compute and graphical applications, it is a fitting choice to create applications to be compiled into SPIR-V.
 
\subsubsection{Vulkan}

Vulkan also satisfies mobile edge computing needs well due to its ground-up design towards multi-threading: parallelizable workloads such as rendering, map-reduction, and machine learning can be done in an efficient manner. Vulkan achieves this by having a static global state, no driver synchronizations, and by separating work generation from work submission. Compared to other \gls{GPU} \glspl{API}, Vulkan is lower-level, making it possible to maximize the performance of both mobile and server hardware. The performance advantages do not come free: the programmer must handle resources, synchronization, memory allocation, and work submission. Also, with Vulkan, error checking, state validation, and shader compilation are separate tools. The tools are left out from application deployments, which reduces system overhead \cite{blackert2016evaluation}.

\subsubsection{SPIR-V}

SPIR-V is a simple binary intermediate language for graphical shaders and compute kernels. Its goals include: (1) providing a target-language for new front ends for novel high-level languages, (2) low-level enough to require a reverse-engineering step to reconstruct source code, and (3) improve portability by enabling shared tools to generate or operate on it. All results of intermediate operations are strictly static single assignment form. In this study, SPIR-V is an instrumental part of the research, as it enables a single, both compute and shader supporting \gls{IR}, to be shared as-is between a \gls{UE} and \gls{MEC} server \cite{spirv}.

\section{Implementation and Results}
\label{ch:ir}


Tests were measured on \gls{COTS} desktop architecture using Nvidia RTX 2080 on Windows 10 and on ARM-architecture using Nvidia Jetson TX2 and Ubuntu 18.04. In both cases, the Vulkan version used was $1.1.97$. We consider the RTX 2080 to demonstrate a \gls{MEC} server hardware and the Jetson TX2 that of \gls{UE}'s. The software implementation for the tests\footnote{Available on GitHub at https://github.com/toldjuuso/haavisto2019gpu} used Rust and Vulkano \cite{vulkano} library, which implements a Vulkan API wrapper. With Vulkano, we used GLSL shaders, which Vulkano compiles to SPIR-V during runtime.

In the first experiment, we measured framebuffer generation times for a simple graphics application, with the results presented in Table \ref{tb:draw}. This is the case of transacting IRs between UE and MEC and hence avoiding video decoding. The baseline shows draw times with video decoding on the UE. We measured that over 1000 executions, the 99th percentile latency on the RTX 2080 was of $2.2$~ms, and the mean latency was $0.39$~ms. On the Jetson, the 99th percentile latency was $1.2$~ms, and the mean latency was $0.60$~ms. We note that the smaller variance on the Jetson might be explained by the \gls{SoC} design of the computer and by software differences between ARM-based Linux and desktop Windows 10.


In the second experiment, we measured cold-start times of a GPU compute kernel, with the results presented in Table \ref{tb:compute}. Here, we did 64k integers multiplication. The idea was to see the latency it takes to start an arbitrary GPU program and copy those results to the CPU.
In specific, we measured the time it takes to dispatch a command buffer to the GPU, and then execute, synchronize, and copy those results back to the CPU. In this regard, we found that for general usage of our proposed approach, a \gls{RPC} protocol should be in place, which orchestrates program loading and buffer preparation. Now, the results indicate the time it takes for an application to be continued after it has been migrated form a UE to a MEC server, vice versa, or between two MEC servers. Assuming this, we measured that over 1000 executions, the 99th percentile latency on the RTX 2080 was $1.4$~ms, and the mean latency was $0.7$~ms. On the ARM-based Jetson, the 99th percentile latency was $4.3$~ms, and the mean latency was $1.8$~ms.



\begin{table} 
\label{tb:draw}
\caption{Results in-light of measuring application cold-start time.}
\begin{center}
\begin{threeparttable}[b]
\begin{tabular}{c |  c|c|c  }
& AVG & SD & 99th \\
\hline
RTX 2080 & 0.7ms & 0.2ms & 1.4ms \\
Jetson TX2 & 1.8ms & 0.5ms & 4.3ms
\end{tabular}
\end{threeparttable}
\end{center}
\end{table}

\begin{table} 
\label{tb:compute}
\caption{Results in-light of edge-graphics use-case, measuring single frame draw times.}
\begin{center}
\begin{threeparttable}[b]
\begin{tabular}{c |  c|c|c  }
& AVG & SD & 99th \\
\hline
Baseline\tnote{a} & 8.3ms & 1.1ms & \\
RTX 2080 & 0.4ms & 0.4ms & 2.2ms \\
Jetson TX2 & 0.6ms & 0.4ms & 1.2ms
\end{tabular}
\begin{tablenotes}
\item [a] Samsung S7, decoding h264 video \cite{kamarainen2017measurement}
\end{tablenotes}
\end{threeparttable}
\end{center}
\end{table}

\section{Discussion}
\label{ch:di}

\begin{figure}
  \centering
  \includegraphics[width=0.9\columnwidth]{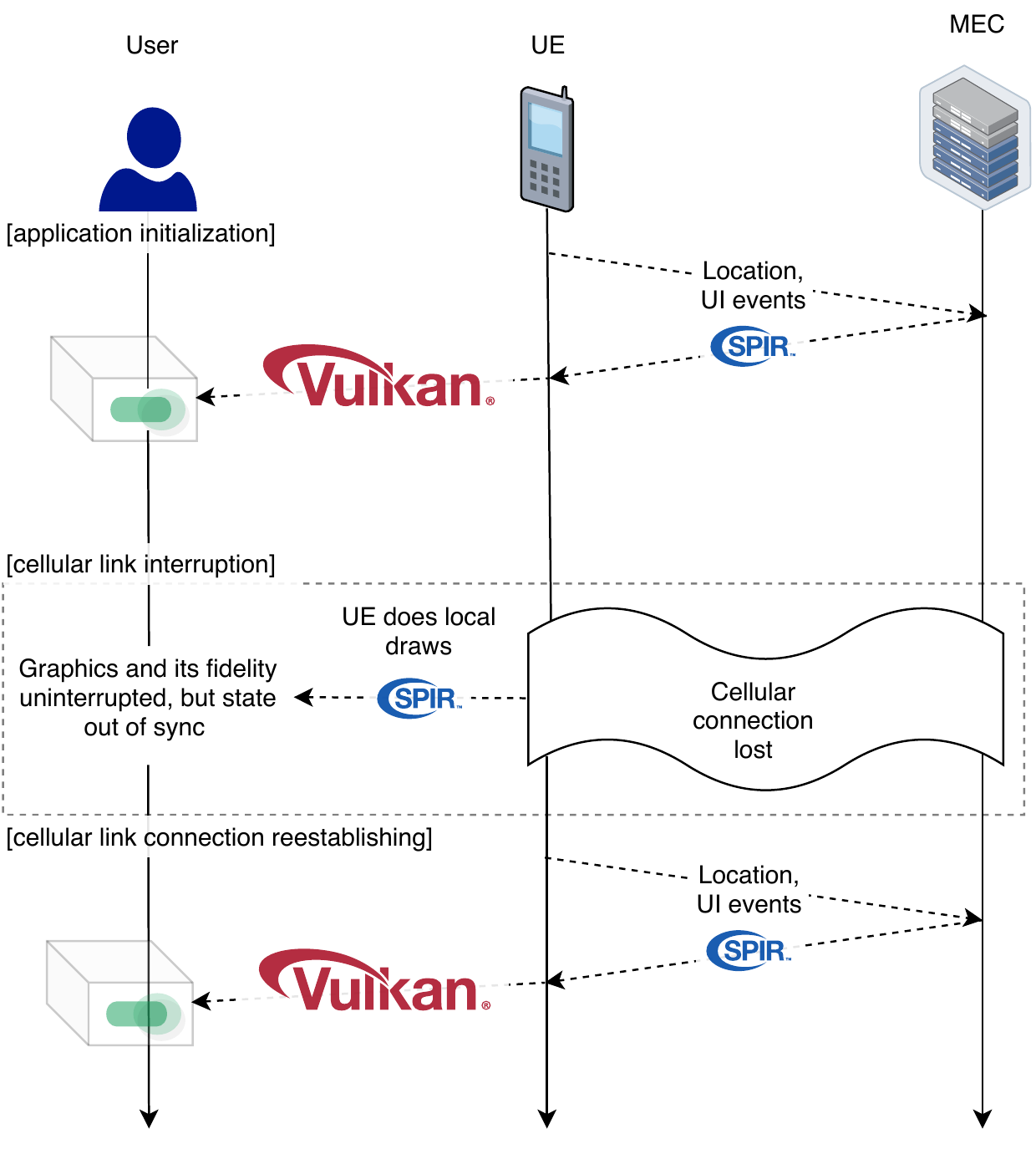}
  \caption{High-level illustration of the RPC protocol.}
  \label{fig:protocol}
\end{figure}

\begin{figure*}
  \centering
  \includegraphics[width=0.9\textwidth]{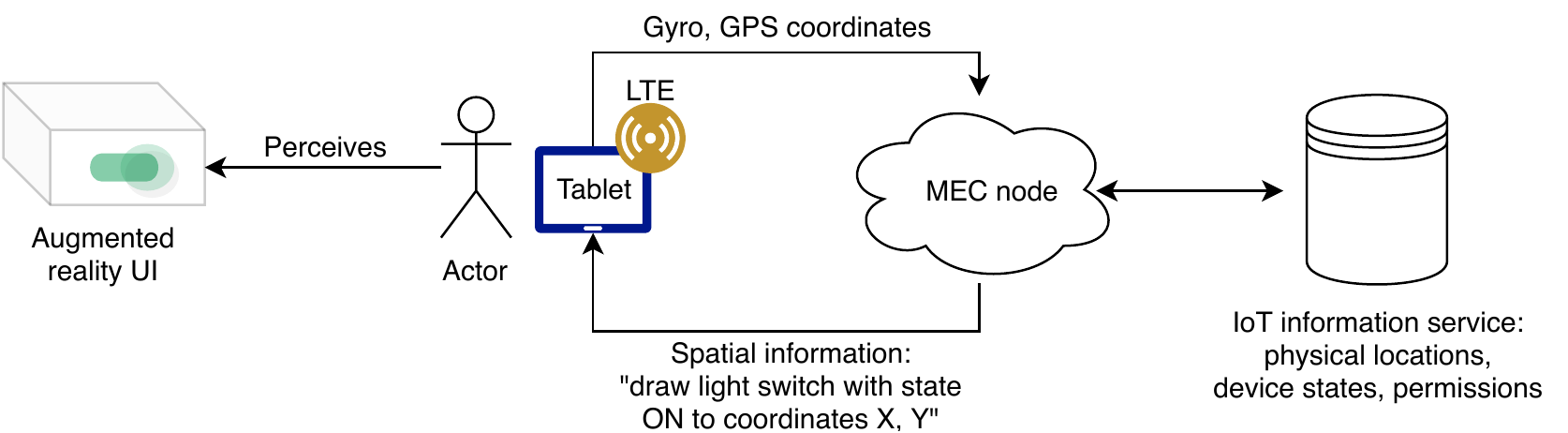}
  \caption{MEC-dependant application working principle.}
  \label{fig:edgeapp}
\end{figure*}

We observed microsecond redraw times for graphics, hence the approach presented in this paper supports running graphics applications at high frame rates. Second, we observed cold-start times of millisecond scale for compute kernels, hence our approach can be used to reduce application migration time from seconds to milliseconds. In our previous study \cite{okwuibe2019orchestrating} we used containers and learnt that it takes seconds to proactively move a container-based session managed by Kubernetes from a MEC server to another.

Generally, we consider direct communication of GPU IRs and the parameters of such, as shown in Fig.~\ref{fig:protocol}, as an interesting approach to reduce device latency. As shown in the figure, this approach can be used to handle network disruptions as well: the UE can compile, produce or store shaders indicating a disconnection to MEC's data services. Until the connection has been re-established, the UE could show the latest but out-of-sync data on-screen, and thus provide degraded yet functional application experience to the user. Regarding rendering in general, compared to a video-based approach \cite{kamarainen2017measurement}, per the working principle of Fig.~\ref{fig:protocol}, such an approach eliminates the need to spend $8.3$~ms in video decoding. Secondly, on bandwidth, irrespective of the resolution used on the \gls{UE}, the SPIR-V \gls{IR} representation remains constant size. This is because the pixels density can be left to be decided (and thus be dependent) by the \gls{UE}. Only the location relative to the screen bounds, and the geometry of the graphics is communicated, not its fidelity. In general, shaders to draw basic shapes, like rectangles, takes only kilobytes of data with SPIR-V. Our insight is that basic shapes could be enough for basic user-interfaces for \gls{MR} applications, such as what we used in our previous study in MEC-dependant \gls{MR} UE \cite{haavisto_2019tiot}. That is if complete virtual surroundings need not be created, but instead merging augmented layers with the physical world, then we stipulate that the useful applications could be built following (e.g., Fig. \ref{fig:edgeapp}) which focus on communication with physically close-by humans and devices. Such applications could be, e.g., smart living environments, where switches made of simple shapes are illuminated if the \gls{UE} is pointing to some physical object.

In addition to the graphics use-case, it might be viable to run certain \glspl{NFV} in an accelerated manner this way. As starting the computing kernels is measured in milliseconds, \gls{NFV} operating this way would have low application migration time since the disk and network-restricted execution environment of \glspl{GPU} might not require virtualization efforts for data protection. Similarly, it might be viable to run certain edge computing applications for \glspl{UE} this way. We envision such applications having a higher quality of service than container-based approaches, due to the low downtime in case the edge application must be migrated to a new node, or possibly taken over by the \gls{UE} in the case of network disruption.

Future research includes distilling the presented \gls{GPU}-based approach into a \gls{DSL}. Because SPIR-V is designed to suit new experimental languages \cite{spirv}, we deem a \gls{DSL} as a workable extension to our work. Such \gls{DSL} could focus on addressing the unique challenges of either the edge-graphics-applications or the general computation problems for \gls{NFV}.

\section{Conclusion}
\label{ch:co}

In this study, we considered the feasibility and technical challenges of using interoperable \gls{GPU} kernels to reduce latency in \gls{MEC} of \gls{5G} cellular networks. We conclude that software-wise, prerequisites for nurturing the paradigm exists. However, software rearchitecting of current pipelines is a must, and expert knowledge in \gls{GPU} domain-specific languages and architecture is required. Yet, compared to previous studies, our approach reduces \gls{UE} latency for graphical applications by sevenfold. For \gls{NFV} applications, a GPU application migration between nodes could be done in $1.4$~ms. Overall, we consider the approach having the capability to speed the end-to-end edge computing pipeline for use-cases requiring graphics, especially those in \gls{MR}. 

\section{Acknowledgements}

This research is financially supported by the Academy of Finland 6Genesis Flagship (grant 318927) and by the AI Enhanced Mobile Edge Computing project, funded by the Future Makers program of Jane and Aatos Erkko Foundation and Technology Industries of Finland Centennial Foundation. Thanks to Jani Saloranta for reading drafts of this paper.

\bibliographystyle{ieeetr}
\bibliography{main}
\end{document}